\documentclass{article}

\usepackage{epsfig} 
\newlength{\abstractwidth} 
\renewcommand{\thefootnote}{\fnsymbol{footnote}} 
\renewcommand{\thanks}[1]{\footnote{#1}} 
\newcommand{\starttext}{ 
\setcounter{footnote}{0} 
\renewcommand{\thefootnote}{\arabic{footnote}}} 
\renewcommand{\theequation}{\thesection.\arabic{equation}} 
\newcommand{\bea}{\begin{eqnarray}} 
\newcommand{\eea}{\end{eqnarray}} 
\newcommand{\beq}[1]{\begin{equation} \label{#1}} 
\newcommand{\be}{\begin{equation}} 
\newcommand{\ee}{\end{equation}}

\def\12{{1 \over 2}} 
\newcommand{\half}{{1\over 2}}

\begin{document} 
\renewcommand{\theequation}{\thesection.\arabic{equation}} 
\bigskip
\centerline{\Large \bf {An Alternative Approach to Unification of Gauge and Geometric Interactions}}
\bigskip
\begin{center} 
{\large James Lindesay\footnote{ 
Address: Department of Physics, Howard University, Washington, DC 20059, 
email: jlindesay@howard.edu}
} \\
Computational Physics Laboratory \\
Howard University
\end{center}
\bigskip\bigskip 
\begin{abstract} 

The algebra of the generators for infinitesimal transformations of 
the $\Gamma=\half$ representation of causal spinor fields (Dirac fields)
\emph{explicitly} constructs the Minkowski metric \emph{within} the internal group space as a 
consequence of non-vanishing commutation relations between generators that carry
a single space-time index.  This representation is a subgroup of the set of
all of the generators that transform under the group GL(4). 
The sixteen hermitian generators of GL(4) include the three angular momentum spin matrices, a matrix
proportional to the Dirac matrix $\gamma^0$,  and 12 additional matrices that have the same number of degrees
of freedom as SU(3)$\times$SU(2)$\times$U(1).  In this paper, 
the construction of linearly independent
internal SU(3) and SU(2) local symmetry groups for the causal spinor fields
is demonstrated to necessarily involve the
CKM mixing of three sets of the SU(3) eigenstates as related to the SU(2) eigenstates.

\end{abstract} 

\starttext \baselineskip=17.63pt \setcounter{footnote}{0} 

\setcounter{equation}{0}
\section{Introduction}
\indent

Perhaps the most fundamental characteristic of the standard model implementation
of fundamental micro-physical interactions is their relation to the gauge symmetry
product group SU(3)$\times$SU(2)$\times$U(1).  Local unitary gauge transformations
generally refer to an invariance of physical measurables under transformations
of the particle fields of the form
$\mathbf{\Psi}(x) \rightarrow \tilde{\mathbf{\Psi}}(x) = \mathbf{U}_{(a)}(\underline{\alpha}(x)) \mathbf{\Psi}(x)$,
where $\mathbf{U}_{(a)}$ is a unitary representation of a group of transformations $\mathcal{G}_{(a)}$,
and the group parameters $\underline{\alpha}=\underline{\alpha}(x)$ can take on an arbitrary functional dependency
on the space-time location $x$.  In order to implement this symmetry, vector gauge potentials $A_\mu^{s}(x)$
are introduced using ``minimal coupling", with the requirement that they transform under the gauge transformation
according to
$ \mathbf{U}_{(a)}^{-1}(\underline{\alpha})  {\hbar \over i} { \partial \over \partial x^\mu} \mathbf{U}_{(a)}(\underline{\alpha}) =
 {q_a \over c} \sum_s A_\mu^{s}  \mathbf{U}_{(a)}^{-1}(\underline{\alpha}) \mathbf{G}^{(a)}_s \mathbf{U}_{(a)}(\underline{\alpha})$, where
$ \mathbf{G}^{(a)}_s$ is the generator of infinitesimal translations of group parameter $\alpha^s$. 
The parameter $q_a$ is the charge coupling the gauge fields $A_\mu^{s}(x)$ to particle currents, making these vector
gauge potentials the carriers of micro-physical interactions between the particles.

General relativity has been consistently successful in describing the macro-physical phenomena of gravitation,
as well as the gravitation of coherent quantum states\cite{PDG2012, Overhauser,MullerPetersChu}.
The theory formulates equations of motion for systems on curvilinear space-times, where the
local curvature is generated by the local energy-momentum density.  All measurables can be derived from
a local and symmetric space-time metric function $g_{\mu \nu}(x)$ that through the principle of equivalence can
be transformed into a locally flat Minkowski metric form $\eta_{\alpha \beta}= {\partial x^\mu \over \partial \xi^\alpha}
g_{\mu \nu} {\partial x^\nu \over \partial \xi^\beta}$ on coordinates $\xi (x)$. 
The metric $\eta_{\alpha \beta}$ cannot be explicitly generated by
the Lorentz group algebra through a Casimir-like construction,
since there are no non-abelian generators in that algebra that carry a single space-time index.

The Dirac equation\cite{Dirac}  for spin $\half$ fermions utilizes a matrix algebra to construct a 
relativistic equation for particle wave functions $\psi$ that is linear in the quantum operators for 4-momentum,
generates a positive semi-definite probability density $\psi^\dagger \psi$,
and maintains the expected non-relativistic correspondence with the Schrodinger equation\cite{BjDrell}. 
In particular, the evolution dynamics describing equations that are linear in energy-momentum operators
 is mathematically straightforward\cite{JLFQG}. 
The development of Dirac's equation required the introduction of
4$\times$4 matrices that each carry a single space-time index.

The Dirac formulation  can be extended by developing general operators $\hat{\Gamma}^\mu$
 that require the form $\hat{\Gamma}^\mu \: \hat{P}_\mu$ to be a Lorentz scalar with an eigenvalue linear in the particle mass. 
The field equation that results takes the form
\be
\mathbf{\Gamma}^\mu \cdot \left ( {\hbar \over i} { \partial \over \partial x^\mu} - {q_a \over c} \sum_s A_\mu^{s} \mathbf{G}_s  \right )\,
\hat{\mathbf{\Psi}}_{\gamma}^{(\Gamma)} (\vec{x}) = -\gamma \, m c  \, \hat{\mathbf{\Psi}}_{\gamma}^{(\Gamma)}(\vec{x}) ,
\label{LinearConfigurationSpinorFieldEqn}
\ee
where $m$ is positive semi-definite for all particle types (both particles and antiparticles), and the $\mathbf{\Gamma}^\mu$ are
finite dimensional matrix representations of the operators $\hat{\Gamma}^\mu$.  In this equation, $\gamma$ is
the eigenvalue of the operator  $\hat{\Gamma}^0$ on the standard (rest) state particle spinor, and this eigenvalue takes differing signs for
particles vs. antiparticles.  Using this formulation, energies are always non-negative. 
This form is seen to be invariant under local gauge transformations of the form previously introduced. 
The Dirac matrices result from the
$\Gamma=\half$ finite dimensional representation of the operators  $\hat{\Gamma}^\mu$.
A representation for the $\Gamma=1$ matrices can be found in Appendix D.2.1 of
reference \cite{JLFQG}. 
In what follows, the group structure of the algebra will be briefly discussed, and internal symmetries of the
causal fields $\hat{\mathbf{\Psi}}_{\gamma}^{(\half)}$ will be explored.

\section{Properties of a Closed Group Inclusive of $\mathbf{\Gamma}^\mu$} 
\indent 

The closed algebra that includes the operators $\hat{\Gamma}^\mu$ add the following commutation relations
to those of the Lorentz group:
\be
\begin{array}{lll}
\left [ \Gamma^0 \, , \, \Gamma^k \right] \: = \: {i \over \hbar} \, K_k  , &
\left [ \Gamma^0 \, , \, J_k \right] \: = \: 0  , &
\left [ \Gamma^0 \, , \, K_k \right] \: = \: -i  \hbar \,  \Gamma^k  , \\
\left [ \Gamma^j \, , \, \Gamma^k \right] \: = \: -{ i \over \hbar} \, \epsilon_{j k m} \, J_m  , &
\left [ \Gamma^j \, , \, J_k \right] \: = \: i \hbar \, \epsilon_{j k m} \, \Gamma^m  , &
\left [ \Gamma^j \, , \, K_k \right] \: = \: -i \hbar \, \delta_{j k} \, \Gamma^0  .
\end{array}
\label{ExtLorentzGroupEqns}
\ee
Eigenvalues $\Gamma$ of the Casimir operator
$C_\Gamma \: = \: {1 \over 6} \left ( \,  ( \underline{J} \cdot \underline{J} \,-\, \underline{K} \cdot \underline{K} )/\hbar^2
\,+\, \Gamma^0 \, \Gamma^0 \,-\, \underline{\Gamma} \cdot \underline{\Gamma} \right )$
will label the irreducible representations of this group.
Additional labels can be attributed to the
eigenvalues of the mutually commuting operators $C_\Gamma$,  $\Gamma^0$, $J^2$, and  $J_3$,
given by ${2 \Gamma (\Gamma + 2)  \over 6}, \: \gamma, \: J(J+1) \hbar^2,$, and $M \hbar$, respectively.

Finite dimensional representations of the algebra  have dimension
$N_\Gamma \: = \: {1 \over 3} (\Gamma + 1) (2 \Gamma + 1) (2 \Gamma + 3)$,
and general representation eigenstates $\chi_{\gamma , M} ^{\Gamma , J}$ can be constructed using the spinor form
\be
\chi_{\gamma , M} ^{\Gamma , J}  = 
\sqrt{{(J-M)! \over  (2J)! \: (J+M)! }} \,[x-y]^{\Gamma - J} \, 
\chi_+ ^{(+) \, M+\gamma} \, \chi_+ ^{(-) \, M-\gamma} \,
 \left . 
\left[ {\partial \over \partial x} \,+\, {\partial \over \partial y}   \right] ^{J+M}
x^{J-\gamma} \, y^{J+\gamma} \right | _{
\begin{tiny}
\begin{array}{l}
x=\chi_+ ^{(+)} \, \chi_- ^{(-)} \\ y=\chi_- ^{(+)} \, \chi_+ ^{(-)} 
\end{array}
\end{tiny} } ,
\ee
where the $\chi_\pm^{(\pm)}$ are $\Gamma=\half$ representation spinors with
$(\gamma)=(\pm \half)$ and $M=\pm \half$. 
The signatures (half-integral vs integral) of $\Gamma$, $J$, $\gamma$, and $M$ are necessarily the same, with $0 \le J \le \Gamma$,
$-J \le \gamma \le J$, and $-J \le M \le J$. 
The spinor forms of the operators satisfying the commutation relations (\ref{ExtLorentzGroupEqns})
can be found in \cite{JLFQG}.
The matrices corresponding to $\Gamma={1 \over 2}$ 
 have dimensionality $N_{1 \over 2}=4$,
with $\mathbf{\Gamma^0}=\half \gamma^0_{Dirac}$ and $\mathbf{J}_3$ both diagonal.
This representation  forms a subgroup of GL(4).

\subsection{Development of a group metric on space-time indexes}

For a general group algebra
$\left [ \hat{G}_r \, , \, \hat{G}_s \right ] \: = \: -i \, \sum_m \left ( c_s \right ) _r ^m \, \hat{G}_m $
with appropriately non-vanishing structure constants $ c_{s r}{}^m=- c_{r s}{}^m$,
the Jacobi identity defines a representation in terms of the structure constants which can
be used to construct a group metric $\eta_{a b}$\cite{Hamermesh},
\be
\eta_{a b} \: \equiv \: \sum_{s \, r} \left ( c_a \right )_r ^s \, \left ( c_b \right )_s ^r, \:
\eta^{ab}\equiv  ((\eta)^{-1})_{ab} .
\ee
This \emph{group} metric on the generators defines invariants on products of group generators, such as the 
Casimir operator $\hat{C}_\mathcal{G} \equiv \sum_{r s} \eta^{G_r G_s} \hat{G}_r \hat{G}_s$.
The non-commuting operators $\hat{\Gamma}^\mu$ that carry a space-time index
conjugate to the 4-momentum define a group metric of Lorentz sub-group invariants given by
\be
\eta^{\Gamma^\mu \, \Gamma^\nu} \: = \: -{1 \over 6} \, \eta_{\mu \, \nu}
\ee
where $\eta_{\mu \, \nu}$ is the usual Minkowski metric. 
Thus, the Minkowski metric defining invariant products of group generators
is \emph{explicitly} generated within this closed algebra,
beyond the Lorentz invariance \emph{implicit} in Lorentz transformations. 
This \emph{group theoretic} metric can be used to develop Lorentz invariants of
\emph{any} operators carrying the group indexes of  $\Gamma^\mu$. 
Once space-time translations are incorporated,
since the generators of space-time translations $P_\mu$ transform as covariant 4-vectors under arbitrary coordinate transformations
(of which group transformations are a special subset\cite{JLLSF13}), this group structure 
is explicitly tied to curvilinear space-time dynamics expressing the principle of equivalence.

\section{Construction of Local Invariance Groups}
\indent

In this section, the invariance groups of $\Gamma=\half$ representation eigenstates will be explored.

\subsection{A complete set of linearly independent hermitian generators in GL(4)}

As previously mentioned, the $\Gamma=\half$ representation is a subgroup of GL(4).  There are
16 hermitian matrices in GL(4), including that associated with the U(1) transformation proportional to the identity. 
The subgroup connected to particle representations already includes the 4 hermitian generators $\Gamma^0$ and the angular momenta $J_k$. 
This leaves an additional 11 matrices, which will be constructed as
$\mathbf{T}_j = i \,  \mathbf{\Gamma}^j$ and 
$\mathbf{T}_{j+3} = i \, \mathbf{K}_j$ for $j=1\rightarrow3$,
 two matrices given by
\be
\mathbf{T}_{7} ={i \over 2} \left (
\begin{array}{cc}
\mathbf{0} & \mathbf{1} \\
-\mathbf{1} & \mathbf{0}
\end{array}
\right )
\quad , \quad
\mathbf{T}_{8} ={1 \over 2} \left (
\begin{array}{cc}
\mathbf{0} & \mathbf{1} \\
\mathbf{1} & \mathbf{0}
\end{array}
\right )  \, ,
\ee
where $\mathbf{1}$ and $\mathbf{0}$ are
the $2 \times 2$ identity and zero matrices respectively,
and a final set of
three generators $\mathbf{T}_9, \, \mathbf{T}_{10},$ 
and $\mathbf{T}_{11}$ forming a closed representation of SU(2) on the lower components given by
\be
\mathbf{T}_{j+8} ={1 \over 2} \left (
\begin{array}{cc}
\mathbf{0} & \mathbf{0} \\
\mathbf{0} & \mathbf{\sigma}_j
\end{array}
\right ) ,
\ee
where the $\sigma_j$ are Pauli spin matrices.

It is intriguing that the number of additional hermitian generators in GL(4) is precisely the same as the
number of hermitian generators in $SU(3)\times SU(2) \times U(1)$.  However, the set 
 of 8 Hermitian generators $\mathbf{T}_s$ for $s:1 \rightarrow 8$
do not form a closed algebra.

\subsection{Transformation properties of causal spinor fields}

Causal spinor fields $\hat{\Psi}(x)$ either commute or anti-commute $\left [ \hat{\Psi}(x) , \hat{\Psi}(y) \right ]_\mp =0$
for space-like separations of the space-time coordinates ($\vec{y}-\vec{x}$) of those fields according to
whether the spin is integral or half-integral.
Thus, microscopic causality compels a well defined \emph{local} relationship between components of spinor fields
in configuration space.
The form of a causal linear spinor field that has
the expected properties under parity, time reversal, and charge conjugation is given by\cite{JLFQG}
\bea
\hat{\mathbf{\Psi}}_{\gamma}^{(\Gamma)}(\vec{x}) \equiv
\sum_{J, s_z } \int {m c^2 \, d^3 p \over \epsilon(\mathbf{p})} 
\left [  {e^{{i \over \hbar} ( \mathbf{p} \cdot \mathbf{x} - 
\epsilon (\mathbf{p}) \, t ) } \over
(2 \pi \hbar)^{3/2}} \, \mathbf{u}_{\gamma}^{(\Gamma)}(\vec{p},m,J,s_z)  \,
\hat{a}_{\gamma}^{(\Gamma)}(\vec{p},m,J,s_z)  + \right . \nonumber   \quad  \\ \left .
(-)^{J + s_z} \, {e^{-{i \over \hbar} ( \mathbf{p} \cdot \mathbf{x} - 
\epsilon (\mathbf{p}) \, t ) } \over
(2 \pi \hbar)^{3/2}} \, \mathbf{u}_{-\gamma}^{(\Gamma)}(\vec{p},m,J,-s_z)  \,
\hat{a}_{-\gamma}^{(\Gamma) \dagger}(\vec{p},m,J,s_z)
\right ] ,  \quad
\label{FinalCausalSpinorField}
\eea
where the normalization has been chosen to have non-relativistic correspondence,
${m c^2 \, d^3 p \over \epsilon(\mathbf{p})} \rightarrow d^3 p$ for $p << m c$,
and $s_z$ specifies the z-component of internal angular
momentum to avoid confusion with the mass $m$.
In this expression, the  $\hat{a}_{\gamma}^{(\Gamma)}(\vec{p},m,J,s_z)$ are annihilation operators
of particle states of the given quantum numbers, $\epsilon(\mathbf{p})=\sqrt{|\mathbf{p}|^2 c^2 + m^2 c^4}$, and the
spinors satisfy $\mathbf{\Gamma}^\mu p_\mu \mathbf{u}_{\gamma}^{(\Gamma)}(\vec{p},m,J,s_z) = 
-\gamma \, m c \, \mathbf{u}_{\gamma}^{(\Gamma)}(\vec{p},m,J,s_z)$. 
In this paper, the focus will be on the $\Gamma=\half$ representation, for which the
$\mathbf{u}_{\gamma}^{(\half)}$ have 4 spinor components. 
For a more complete treatment of the algebra, symmetries, and causality properties
of causal spinor fields, the reader is invited to examine sections 4.3 and 4.4 in
reference \cite{JLFQG}.

Since the spinor components of a causal 4-spinor field
$\mathbf{\Psi}^{(\half)}$ maintain local spatial relationships
described in (\ref{FinalCausalSpinorField}),
a particular local Euclidean rotation $\mathbf{R}_{E}(x)$ will be constructed
to define a spinor with a single component.
If the components of the causal 4-spinor are written in the form
\be
\mathbf{\Psi}(x) = \left (
\begin{array}{l}
\phi_1 (x) e^{i \omega_1 (x)} \\
\phi_2 (x) e^{i \omega_2 (x)} \\
\phi_3 (x) e^{i \omega_3 (x)} \\
\phi_4 (x) e^{i \omega_4 (x)} 
\end{array}
\right ) \, \equiv \,  \left (
\begin{array}{l}
\: \: \:  \phi_1 (x) e^{i \omega_1 (x)} \\
-\phi_1 (x) \tan(\zeta_{12}) \sec(\zeta_{13}) \sec(\zeta_{14})  e^{i \omega_2 (x)}\\
-\phi_1 (x) \tan(\zeta_{13})  \sec(\zeta_{14})  e^{i \omega_3 (x)}\\
-\phi_1 (x)  \tan(\zeta_{14})  e^{i \omega_4 (x)}
\end{array}
\right ),
\ee
then the causal spinor can be expressed by the action of a Euclidean rotation
on a single component spinor as follows:
\begin{displaymath}
\mathbf{R}_{14} = \left (
\begin{array}{cccc}
\cos(\zeta_{14}) & 0 & 0 & \sin(\zeta_{14}) e^{ i \omega_{14}} \\
0 & 1 & 0 & 0 \\
0 & 0 & 1 & 0  \\
-\sin(\zeta_{14}) e^{ -i \omega_{14}} & 0 & 0 & \cos(\zeta_{14}) 
\end{array}
\right ) , \quad
\mathbf{R}_{13} = \left (
\begin{array}{cccc}
\cos(\zeta_{13}) & 0 & \sin(\zeta_{13})  e^{ i \omega_{13}} & 0 \\
0 & 1 & 0 & 0 \\
- \sin(\zeta_{13})e^{ -i \omega_{13}} & 0 &  \cos(\zeta_{13}) & 0  \\
0 & 0 & 0 & 1 
\end{array}
\right ) , 
\end{displaymath}
\be
\mathbf{R}_{12} = \left (
\begin{array}{cccc}
\cos(\zeta_{12})  & \sin(\zeta_{12})  e^{ i \omega_{12}} & 0 & 0 \\
 -\sin(\zeta_{12})  e^{- i \omega_{12}} &  \cos(\zeta_{12})   & 0 &0 \\
0 & 0 & 1 & 0 \\
0 & 0 & 0 & 1
\end{array}
\right ) , \,  \omega_{1s} \equiv \omega_1-\omega_s , \,
\mathbf{R}_{E} \equiv \mathbf{R}_{14} \mathbf{R}_{13}\mathbf{R}_{12} , \,
\mathbf{\Psi}(x) =\mathbf{R}_{E}(x) \mathbf{\bar{\Phi}}(x).
\ee
The single component spinor takes the form
\be
\mathbf{\bar{\Phi}}(x) = \left (
\begin{array}{c}
 \sqrt{(\phi_1(x))^2 + (\phi_2 (x))^2 + (\phi_3 (x))^2 + (\phi_4 (x))^2 }\:  e^{i \omega_1 (x)}\\
0 \\
0 \\
0
\end{array}
\right ) = \mathbf{R}_{E}^{-1}(x) \mathbf{\Psi}(x) .
\label{Eq:psibarfromPsi}
\ee
It is important to note that this transformation must be made uniquely at each location $x$. 
This imposes a  \emph{local} nature to any internal symmetries that utilize this construction.

\subsection{SU(2) and SU(3) invariance transformations on $\mathbf{\bar{\Phi}}(x)$}

The previously defined generators $\tau_j \equiv \mathbf{T}_{j+8}$ with $j:1 \rightarrow 3$
form a closed matrix group algebra
$\mathbf{M}^{(2)}$
that transform components in the subspace $\mathbf{S}^{(2)}$:
\be
\mathbf{M}^{(2)}(\underline{\theta})= \left (
\begin{array}{cc}
\mathbf{1} & \mathbf{0} \\
\mathbf{0} & \mathbf{S}^{(2)}(\underline{\theta})
\end{array}
\right ) 
= e^{i \sum \theta^s \tau_s} .
\label{Eq:M2}
\ee
In this expression, the matrix $\mathbf{S}^{(2)}$ is a unitary unimodular transformation matrix in SU(2), and $\mathbf{1}$ is the 2$\times$2 identity matrix. 
Since $\tau_j  \mathbf{\bar{\Phi}}(x) = 0$, the field $ \mathbf{\bar{\Phi}}(x)$ from Eq. \ref{Eq:psibarfromPsi} 
is invariant under transformations involving $ \mathbf{M}^{(2)}$. 
The generators $\tau_j$ share this basis simultaneously with $\mathbf{\Gamma}^0$ and $\mathbf{J}_3$.  

As was previously mentioned, the remaining hermitian
generators in this basis $\{ \mathbf{T}_1,...,\mathbf{T}_8 \}$ do not form a closed algebra.
However, a basis of SU(3) states \emph{can} be constructed through mixing of the hermitian generators in this basis.  This mixing
between precisely three sets of the basis states will next be demonstrated.

Consider the set of SU(3) transformations of the form
\be
\mathbf{M}^{(3)}(\underline{\alpha})= \left (
\begin{array}{cc}
1 &\overline{\mathbf{0}}^T \\
\overline{\mathbf{0}} & \mathbf{S}^{(3)} (\underline{\alpha})
\end{array}
\right ) = e^{i \sum \alpha^b \mathbf{t}_b},
\label{EqM3}
\ee
where $\mathbf{S}^{(3)}$ is a unitary unimodular transformation matrix in SU(3), and $\overline{\mathbf{0}}$ is a 1$\times$3 zero vector. 
The set of eight generators $\{ \mathbf{t}_1, ... , \mathbf{t}_8 \}$ form a closed algebra of SU(3). 
Since $\mathbf{t}_b \mathbf{\bar{\Phi}}(x) =0$, the field $\mathbf{\bar{\Phi}}(x)$ is invariant under transformations involving 
$ \mathbf{M}^{(3)}$.  The SU(3) eigenstates will be defined using this basis.  A set of SU(3) generators
(including 2 diagonal generators) can be found in \cite{JLUGI16}.

The construction of a representation of SU(3) that is inclusive of the hermitian generators
$\{ \mathbf{\Gamma}^0,\mathbf{J}_1,\mathbf{J}_2,\mathbf{J}_3 \}$
in GL(4) necessarily involves the mixing of the three basis states of the transformation (\ref{EqM3}). 
A convenient mechanism for mixing three eigenstates 
is provided through general CKM matrices\cite{Cabibbo,KMmix} embedded within GL(4).
The particular choice for mixing will be 
the set of all transformations on the $3 \times 3$ subspace that leaves the second spin component of
$\mathbf{\bar{\Phi}}(x)$ invariant as demonstrated  below:
\begin{displaymath}
\mathbf{U}_{23} = \left (
\begin{array}{cccc}
1 & 0 & 0 & 0 \\
0 & 1 & 0 & 0  \\
0 & 0 &  \cos(\theta_{23}) & \sin(\theta_{23})  \\
0 & 0 &  -\sin(\theta_{23}) & \cos(\theta_{23}) 
\end{array}
\right ) , \quad
\mathbf{U}_{31} = \left (
\begin{array}{cccc}
\cos(\theta_{31}) & 0 & 0 &  \sin(\theta_{31})e^{- i \delta_{31}}  \\
0 & 1 & 0 & 0 \\
0 & 0 & 1 & 0 \\
- \sin(\theta_{31})e^{i \delta_{31}} & 0 & 0 &  \cos(\theta_{31})   \\
\end{array}
\right ) , 
\end{displaymath}
\be
\mathbf{U}_{12} = \left (
\begin{array}{cccc}
\cos(\theta_{12})  & 0 & \sin(\theta_{12}) & 0  \\
0 & 1 & 0 & 0 \\
 -\sin(\theta_{12}) & 0 &  \cos(\theta_{12})   & 0 \\
0 & 0 & 0 & 1
\end{array}
\right ) , \quad
\mathbf{U}_{CKM} \equiv \mathbf{U}_{23} \mathbf{U}_{31}\mathbf{U}_{12} .
\ee
Most CKM transformations of this type on the generators $\tilde{\mathbf{t}}_s \equiv \mathbf{U}_{CKM} \mathbf{t}_s \mathbf{U}_{CKM}^{-1}$
will produce a set of transformed generators $\tilde{\mathbf{t}}_s$ that both satisfy the algebra of SU(3) as well as can be included among
a set of 11 linearly independent traceless hermitian generators that complete the
set $\{ \mathbf{\Gamma}^0,\mathbf{J}_1,\mathbf{J}_2,\mathbf{J}_3 \}$.

For clarity, consider an example CKM transformation from the SU(3) eigenbasis to that including
the generators $\left \{ \mathbf{T}_1, ... , \mathbf{T}_{11}  \right  \}$
parameterized using data consistent with known mixing between the generations in quark phenomenology, 
($\theta_{12}\rightarrow 0.227, \: \theta_{23}\rightarrow 0.446, \: \theta_{31}\rightarrow 0.0035, \: \delta_{31} \rightarrow 1.96$):
\be
\mathbf{U}_{CKM} = \left (
\begin{array}{cccc}
0.974 & 0 &0.225   &  -0.0013-0.0032 i \\
0 & 1 & 0 & 0 \\
-0.225 +0.00014 i & 0 & 0.973 + 0.000033 i  & -0.0446 \\
-0.00875 - 0.00316 i & 0 & 0.0437 - 0.00073 i &  0.999 
\end{array}
\right )  .
\ee
The set of 15 linearly independent hermitian generators
$\left \{  \mathbf{\Gamma}^0, \mathbf{J}_1, \mathbf{J}_2, \mathbf{J}_3, \mathbf{T}_1, ... , \mathbf{T}_{11}  \right  \}$
can be directly decomposed in terms of an alternative set of 15
linearly independent hermitian generators given by
\begin{displaymath}
\left \{  \mathbf{\Gamma}^0, \mathbf{J}_1, \mathbf{J}_2, \mathbf{J}_3, \tilde{\mathbf{t}}_1, ... , \tilde{\mathbf{t}}_8 ,
 {\mathbf{\Delta}}_1 \equiv (\mathbf{T}_1-\mathbf{T}_5)/2, 
{\mathbf{\Delta}}_2  \equiv (\mathbf{T}_2+\mathbf{T}_4)/2, 
{\mathbf{\Delta}}_3  \equiv (\mathbf{T}_3+\mathbf{T}_7)/2  \right \}.
\end{displaymath}
Although the set of matrices $\{ \tilde{\mathbf{t}}_1, ... , \tilde{\mathbf{t}}_8  \}$ now generate a closed
SU(3) algebra in this basis,
the remaining generators ${\mathbf{\Delta}}_1 , \, {\mathbf{\Delta}}_2 , $ and ${\mathbf{\Delta}}_3 $
\emph{do not} form a closed algebra.  The transformed SU(3) generators $\tilde{\mathbf{t}}_b$
are decomposed in (\ref{Eq:CKMSU3}).
\be
\begin{scriptsize}
\begin{array}{cccccccccccccccc}
 &  \Gamma^0 &  J_1 &  J_2  &  J_3 & T_{1} & T_{2} & T_{3} & T_{4} & T_{5} & T_{6} & T_{7} & T_{8} & T_{9} & T_{10} & T_{11} \\
\tilde{\mathbf{t}}_1  &  -.295 & 0 & 0 &  - .295 & - .071 &  112 & - 1.58 &  112 & .071 & - 5.65  & - 1.58 &  - 5.65 &  970 & 0 &  - 86.8 \\
\tilde{\mathbf{t}}_2  &  -.730 & 0 & 0 & - .730 & 112 & - .071 & - 4.38 & - .071 & - 112 & - 1.58 & - 4.38 & - 1.58 & - .065 & - 974  &  .003 \\
\tilde{\mathbf{t}}_3  &  -25.3 & 0 & 0 & - 25.3 &  - 1.7 &  - 5.58 & .076 & - 5.58 & 1.7 & - 109 & .076 & -109 & - 87.1 & .711 & - 945  \\
\tilde{\mathbf{t}}_4  & 0 & -1.31 &  3.24  & 0 & 0 & 22.3 & 0 &  -22.3 & 0 &  - 499 & 0 & 499 & 1.31 & - 3.24 & 0 \\
\tilde{\mathbf{t}}_5 & 0 &  -3.24 & - 1.31 & 0 &  - 22.3 & 0 & - 500 & 0 &  - 22.3 & 0 &  500 & 0 & 3.24 & 1.31 & 0 \\
\tilde{\mathbf{t}}_6 & 0 & 225 & 0 & 0 &  - .016 & - 487 & - .365 & 487 & - .016 &  - 21.9 & .365 & 21.9 & - 225 & 0 & 0  \\
\tilde{\mathbf{t}}_7 & 0 & 0 &  225 & 0 & 487 & - .016 & - 21.9 & .016 & 487 & .365 & 21.9 & - .365 & 0 & - 225 & 0 \\ 
\tilde{\mathbf{t}}_8 & -975 & 0 & 0 &  1025 & - 1.54 & 4.26 & .069 & 4.26 & 1.54 & 110 & .069 & 110 & - 1.97 & - .711 & - 1051 
\end{array}
\label{Eq:CKMSU3}
\end{scriptsize}
\ee
For each of the transformed generators $\tilde{\mathbf{t}}_b=\sum_j k_b^j \mathbf{G}_j$,
the coefficients $k_b^j$ are to be obtained by multiplying the number in the column
under the previously defined generator $\mathbf{G}_j$ by $10^{-3}$ in (\ref{Eq:CKMSU3}). 
It is clear that the set of independent generators $ \{  \mathbf{\Gamma}^0, \mathbf{J}_1, \mathbf{J}_2, \mathbf{J}_3, \tilde{\mathbf{t}}_1, ... , \tilde{\mathbf{t}}_8 ,  {\mathbf{\Delta}}_1,{\mathbf{\Delta}}_2, {\mathbf{\Delta}}_3   \}$
that includes a closed SU(3) sub-algebra $\{ \tilde{\mathbf{t}}_1, ... , \tilde{\mathbf{t}}_8  \}$
necessarily mixes in an invertible manner the basis 
$\{  \mathbf{\Gamma}^0, \mathbf{J}_1, \mathbf{J}_2, \mathbf{J}_3,
\mathbf{T}_1, ... , \mathbf{T}_{11} \}$ that includes an internal closed SU(2) sub-algebra $\{\mathbf{T}_{9}, \mathbf{T}_{10} , \mathbf{T}_{11} \}$
on $\mathbf{\bar{\Phi}}(x)$.
The  independent group of SU(3) transformations on the 4-spinors generated by $\{ \tilde{\mathbf{t}}_1, ... , \tilde{\mathbf{t}}_8  \}$
will be denoted $\mathbf{U}^{(3)}(\underline{\alpha} (x))$, so that
\be
\mathbf{M}^{(3)} \equiv \mathbf{U}^{-1}_{CKM} \,  \mathbf{U}^{(3)} \:  \mathbf{U}_{CKM} ,
\label{Eq:M3}
\ee
where $\mathbf{M}^{(3)}$ is of the form defined in Eq. \ref{EqM3}.

\subsection{Local gauge symmetries of spinor fields}

A  local internal SU(3) symmetry on the causal spinor field $\hat{\mathbf{\Psi}}_{\gamma}^{(\half)}(\vec{x})$
from (\ref{FinalCausalSpinorField}) can be constructed
using the internal SU(3) symmetry transformation $\mathbf{M}^{(3)}$ on $\mathbf{\bar{\Phi}}(x)$
from (\ref{Eq:M3}), along with the relationship of the transformed spinor $\mathbf{\bar{\Phi}}(x)$ to the general causal spinor
field expressed in (\ref{Eq:psibarfromPsi}), via 
\be
\mathcal{U}^{(3)} (x) =\mathbf{R}_E (x) \: \mathbf{M}^{(3)}(\underline{\alpha}(x)) \:  \mathbf{R}_E^{-1}(x) = 
\mathbf{R}_E (x)  \mathbf{U}^{-1}_{CKM} \,  \mathbf{U}^{(3)}(\underline{\alpha}(x)) \:  \mathbf{U}_{CKM}  \mathbf{R}_E^{-1}(x)  ,
\label{Eq:causalSU3}
\ee
where $\alpha^s(x)$ are the eight local group parameters of SU(3).
Similarly, the transformation $\mathbf{M}^{(2)}$ from (\ref{Eq:M2}) on 
$\mathbf{\bar{\Phi}}(x)$ defines a local internal SU(2) symmetry on the causal spinor field $\hat{\mathbf{\Psi}}_{\gamma}^{(\half)}(\vec{x})$
under transformations
\be
\mathcal{U}^{(2)} (x) =\mathbf{R}_E (x) \: \mathbf{M}^{(2)}(\underline{\theta}(x) ) \: \mathbf{R}_E^{-1}(x)  ,
\label{Eq:causalSU2}
\ee
where $\theta^j (x)$ are the three local group parameters of SU(2). 
Although both (\ref{Eq:causalSU3}) and (\ref{Eq:causalSU2}) are local internal symmetries on the causal spinor field
$\mathcal{U}(x) \mathbf{\Psi}(x) =\mathbf{\Psi}(x) $,
the eigenbases of the internal SU(2) and SU(3) symmetries are related via
CKM mixing of the three SU(3) eigenstates in the enlarged unified group GL(4) via $\mathbf{U}_{CKM}$. 

Furthermore, the independent sets of hermitian generators  $ \{  \mathbf{\Gamma}^0, \mathbf{J}_1, \mathbf{J}_2, \mathbf{J}_3,
\tilde{\mathbf{t}}_1, ... , \tilde{\mathbf{t}}_8 , {\mathbf{\Delta}}_1, {\mathbf{\Delta}}_2, {\mathbf{\Delta}}_3   \}$
and $\{  \mathbf{\Gamma}^0, \mathbf{J}_1, \mathbf{J}_2, \mathbf{J}_3,
\mathbf{T}_1, ... , \mathbf{T}_{11} \}$ include
local \emph{gauge} invariance transformations under
SU(3) and SU(2) for \emph{interacting} fields satisfying (\ref{LinearConfigurationSpinorFieldEqn})
that connect the two basis sets via CKM mixing, where
 \be
\mathbf{U}^{(3)} (\underline{\alpha}) = e^{i \sum_b \alpha^b \tilde{\mathbf{t}}_b} 
\textnormal{ and }
\mathbf{U}^{(2)} (\underline{\theta}) = e^{i \sum_s \theta^s \tau_s} .
\label{Eq:gaugeSU3SU2}
\ee
In these expressions, $\underline{\alpha}=\underline{\alpha}(x)$ are local group parameters of SU(3) and
$\underline{\theta}=\underline{\theta}(x)$ are local group parameters of SU(2) under gauge transformations of the field
$\mathbf{\Psi}(x) \rightarrow \tilde{\mathbf{\Psi}}(x) = \mathbf{U}^{(3,2)}(\underline{\beta}(x)) \mathbf{\Psi}(x)$. 
The topology of the local mapping of gauge group parameters $\underline{\beta}(x)$ in space-time determines the monopole structure of
the sources of the gauge interactions (see section 4.2 of \cite{JLFQG}), as well as the properties of
any higher dimensional structures (e.g., strings, etc.) in the formulation.

\section{Symmetric Tensors and Conservation Properties}

It is of interest to determine whether a symmetric energy-momentum tensor
suitable for geometrodynamics can be generated from a
Lagrangian that produces the spinor field equations. 
It should be noted that the equations that define the properties
of the spinors given in (\ref{LinearConfigurationSpinorFieldEqn}) generally need not
be the same as the equations used to construct the energy momentum
tensor, and one can \emph{always} construct symmetric energy momentum tensors
from a Lagrangian of the form
\bea
\mathcal{L}=  \half \left \{  {1 \over  m} \left [  
 \left ( {\hbar \over i} { \partial \over \partial x^\mu} - {q_a \over c} \sum_s A_\mu^{s} \mathbf{G}_s  \right ) \mathbf{\Psi} \right ]^\dagger 
g^{\mu \nu}  \left ( {\hbar \over i} { \partial \over \partial x^\nu} - {q_a \over c} \sum_s A_\nu^{s} \mathbf{G}_s  \right )
\mathbf{\Psi}
+  \, m c^2  \,  \mathbf{\Psi}^\dagger  \mathbf{\Psi} 
 \right  \}  +  \nonumber \\
-{1 \over 16 \pi }Tr(\mathbf{F}^{\mu \nu} \mathbf{F}_{\mu \nu})
-{c^4 \over 16 \pi G_N} \mathcal{R},  \quad
\label{Eq:GravLagrangian}
\eea
where $g$ is the determinant of the space-time metric tensor,
$\mathbf{F}_{\mu \nu}=\partial_\mu \mathbf{A}_\nu - \partial_\nu \mathbf{A}_\mu
- i {q_a \over \hbar c}[\mathbf{A}_\mu,\mathbf{A}_\nu]$,
$\mathbf{A}_\mu \equiv  \sum_s A_\mu^{s} \mathbf{G}_s$, and $\mathcal{R}$ is the Ricci scalar. 
Gravitational field equations result from (\ref{Eq:GravLagrangian}) by requiring that
${2 \over \sqrt{-g}} {\delta \over \delta g_{\mu \nu}} \left ( \sqrt{-g} \, \mathcal{L}  \right )=0$. 
Causal fields (\ref{FinalCausalSpinorField})  that satisfy the spinor field equations  (\ref{LinearConfigurationSpinorFieldEqn})
will \emph{also} satisfy the Klein-Gordon equation resulting from this Lagrangian form, which directly produces
symmetric energy-momentum tensors.

To determine if the symmetric, locally conserved tensors can also be constructed from
a Lagrangian form that generates the spinor field equations,
it suffices to examine non-interacting fields.
Spinor field equations  (\ref{LinearConfigurationSpinorFieldEqn}) will result from
a Lagrangian of the form
\be
\mathcal{L}_s={1 \over 2} \overline{\mathbf{\Psi}} \left [  
\mathbf{\Gamma}^\mu \cdot \left ( {\hbar \over i} { \partial \over \partial \xi^\mu}   \right )\,
\mathbf{\Psi} + \gamma \, m c  \, \mathbf{\Psi} 
\right ]  + cc
={1 \over 2} \overline{\mathbf{\Psi}} \left [  
\mathbf{\Gamma}^\mu {\partial x^\beta \over \partial \xi^\mu} \cdot \left ( {\hbar \over i} { \partial \over \partial x^\beta}   \right )\,
\mathbf{\Psi}
+ \gamma \, m c  \, \mathbf{\Psi} 
\right ]  + cc.
\ee
where $\overline{\mathbf{\Psi}}$ is the Dirac conjugate
of field $\mathbf{\Psi}$,  $cc$ refers to the complex conjugate of the previous term,
and $\xi^\mu$ are locally flat coordinates.
A conserved second rank tensor can be constructed consistent with the Euler-Lagrange equations of the form
\be
\mathbf{\Pi}^\mu \equiv {\partial \mathcal{L} \over \partial (\partial_\mu \mathbf{\Psi}) } , \quad
\mathcal{T}^{\mu \nu} = \mathbf{\Pi}^\mu g^{\nu \beta}  \partial_\beta \mathbf{\Psi} + cc
-g^{\mu \nu} \mathcal{L},
\ee
However, in this case the tensor $\mathcal{T}^{\mu \nu}$ is not symmetric.
Following Belinfante\cite{Belinfante}\cite{WeinbergQFT}, the tensor describing
internal transformations under general Lorentz transformations given by
\be
\mathcal{S}^{\beta \mu \nu} = {i \hbar \over 2} \left (
\mathbf{\Pi}^\beta  [\mathbf{\Gamma}^\mu,\mathbf{\Gamma}^\nu] \mathbf{\Psi}
- \mathbf{\Pi}^\mu  [\mathbf{\Gamma}^\beta,\mathbf{\Gamma}^\nu] \mathbf{\Psi}
- \mathbf{\Pi}^\nu  [\mathbf{\Gamma}^\beta,\mathbf{\Gamma}^\mu] \mathbf{\Psi} \right )  + cc
\ee
can be used to construct a symmetric second rank tensor $T^{\nu \mu}=T^{\mu \nu} \equiv
\mathcal{T}^{\mu \nu} + \partial_\beta \mathcal{S}^{\beta \mu \nu}$ that satisfies the conservation equation
$\partial_\mu T^{\mu \nu} =0$ in flat space-time (since $\mathcal{S}^{\beta \mu \nu}$ is antisymmetric
under the interchange $\beta \leftrightarrow \mu$).  However, using curvilinear coordinates, the results of the construction
are as follows:
\be
T^{\mu \nu} \equiv \mathcal{T}^{\mu \nu} + D_\beta \mathcal{S}^{\beta \mu \nu}  ,\quad
D_\mu T^{\mu \nu} = -\half R^{\nu}_{\lambda \beta \mu} \mathcal{S}^{\beta \mu \lambda}.
\ee
Thus, the covariant divergence of the Belinfante construct does not generally vanish in
curved space-times. 
That this construct does not generate an energy-momentum tensor that can drive
the geometrodynamics should not be surprising, since for integer values of $\Gamma$, there are states
satisfying (\ref{LinearConfigurationSpinorFieldEqn}) with finite mass,
yet vanishing eigenvalues of $\mathbf{\Gamma}^\mu \hat{P}_\mu$, for which $\gamma=0$.  These states must
nonetheless carry non-vanishing energy-momentum.

\section{Conclusions}
\indent

The fundamental representation of causal spinor fields has been shown to unify a set of internal local symmetries
including a U(1) symmetry along with 11 additional hermitian generators that can represent
a linearly independent SU(2) symmetry or a linearly independent
SU(3) symmetry, but not both simultaneously.  The eigenbasis of the SU(3) internal symmetry has been
related to that of the SU(2) internal symmetry via a CKM transformation mixing the symmetries in GL(4)
consistent with observed phenomenology.
The closed representation of SU(2) can share the same basis
as that in which $\Gamma^0$ and $J_3$ are diagonal, whereas the SU(3) eigenbasis with
two diagonal generators cannot share this basis. 

Group algebraic invariants \emph{explicitly} generate the Minkowski metric using the non-abelian
algebra of generators that carry single space-time indexes, thereby extending interior group structure to the dynamics
of general coordinate transformations.  The geometrodynamics of general relativity follows directly via
the principle of equivalence. 

On-going efforts examine how the gauge bosons of the internal SU(2)$\times$U(1) pre-symmetry of the
$\Gamma=\half$ representation can
relate to the quanta of the \emph{massive} $\Gamma=1$ boson representation with $N_{\Gamma=1}=10$, which
contains a self-adjoint scalar particle, a self-adjoint vector particle,
another vector particle, and its adjoint vector particle.
The self-adjoint $\gamma=0$ states are degenerate in their eigenvalue, and can freely mix
massive as well as massless vector states in a manner that preserves the
degrees of freedom.


\end{document}